\documentclass[10pt]{aastex63}
\usepackage{amsmath}
\usepackage{graphicx}
\usepackage{epstopdf}
\usepackage{color}

\newcommand{\D}[1]{\, d #1 \,}
\newcommand{\vc}[1]{{\bf#1}}
\newcommand{\vch}[1]{{\bf \hat{#1}}}
\newcommand{\avg}[1]{\left< #1 \right>}

\newcommand{\mcl}[1]{\mathcal{#1}}

\newcommand{\e}[2]{\hat{e}_{{#1}_s{#2}}}
\newcommand{\sqrtd}[1]{\sqrt{d\bf{#1}} \:}

\newcommand{\tpki}{{\bf T}_p(\vch{k}_i)}

\begin{document}

\title{Explicit form of the random field spectral representation and some applications}

\author{A. Chepurnov}

\begin{abstract}
We present here an explicit form of the random spectral measure element, what allows us to express a stationary random field as a stochastic integral explicitly depending on its power spectrum and a spectral tensor if the field is a vector one. It has been shown here that convergence mechanism of such integral is significantly different from the one of the Fourier transform and that the traditional formalism is a partial limiting case of the one presented here. The fact that there is an explicit expression of a random field makes calculation of higher order statistics of it much more straightforward (see for example \citealp{Ch20}). For a vector field such expression contains a projection of an isotropically distributed random vector by a spectral tensor, what makes geometrical interpretation of harmonics behavior possible, simplifying its analysis (see \S\ref{sect:vf}). This spectral representation also makes straightforward numerical generation of a random field, what is extensively used by \cite{Ch20}. We also present here some practical applications of this formalism.
\end{abstract}



\section{Spectral representation of a scalar field} \label{sect:sf}

The following considerations are based on results obtained in \cite{IR70}, \cite{R90} and \cite{Ch98}.

\subsection{Spectral representation}

Let us write the correlation function of a homogeneous random field $\rho(\vc{r})$ through its power spectrum:
\begin{equation} \label{eq:scf}
C(\vc{r}) \equiv \avg{\rho^*(\vc{0})\rho(\vc{r})} = \int e^{i\vc{k}\vc{r}} \mcl{F}(d\vc{k}) = \int e^{i\vc{k}\vc{r}} F^2(\vc{k}) \D{\vc{k}} 
\end{equation}
\noindent where $\mcl{F}(\cdot)$ is a correlation spectral measure and $F^2$ is the correspondent power spectrum.

Every homogeneous field admits a spectral representation as follows:
\begin{equation} \label{eq:ssr_orig}
\rho(\vc{r}) = \int e^{i\vc{k}\vc{r}} \Phi(d\vc{k}) 
\end{equation}
\noindent where $\Phi(\cdotp)$ is a complex random measure in $\mathbb{R}^3$, satisfying\footnote{It follows from Eq. (\ref{eq:sphi}) that such field is Gaussian, see \S \ref{sect:agc}.}
\begin{equation} \label{eq:sphi}
\avg{\Phi(A)\Phi^*(B)} = \mcl{F}(A\cap B) 
\end{equation}

Consequently, measure elements must conform the following symbolic rule:
\begin{equation} \label{eq:sphi_symb_orig}
\avg{\Phi(d\vc{k})\Phi^*(d\vc{k}')} = \delta_{\vc{k}\vc{k}'} \mcl{F}(d\vc{k}) =  \delta_{\vc{k}\vc{k}'} F^2(\vc{k}) \D{\vc{k}}
\end{equation}

We would like to improve the notation Eq. (\ref{eq:ssr_orig}), exposing the internal structure of the spectral measure element. If we introduce a complex random field $\xi$ as follows
\newpage
\begin{equation} \label{eq:sphi_symb}
\Phi(d\vc{k}) = F(\vc{k}) \, \xi(\vc{k}) \sqrtd{k}
\end{equation}
\begin{equation} \label{eq:xi_corr}
\avg{\xi(\vc{k})\xi^*(\vc{k}')} = \delta_{\vc{k}\vc{k}'},
\end{equation}
\noindent we satisfy Eq. (\ref{eq:sphi_symb_orig}). Then, Eq. (\ref{eq:ssr_orig}) can be rewritten in the following form:
\begin{equation} \label{eq:ssr}
\rho(\vc{r}) = \int e^{i\vc{k}\vc{r}} F(\vc{k}) \, \xi(\vc{k}) \sqrtd{k} 
\end{equation}
\noindent where $\xi(\vc{k})$ conforms Eq. (\ref{eq:xi_corr}) and $F(\vc{k})$ is square root of the power spectrum. In addition, if $\rho \in \mathbb{R}$,
\begin{equation} \label{eq:sxi}
\xi(-\vc{k}) = \xi^*(\vc{k}).
\end{equation}


\subsection{Delta-symbol with continuous arguments}

It should be noted that the function $\delta_{\vc{k}\vc{k}'}$ involved in the correlation of the random field $\xi$ (Eqs. \ref{eq:xi_corr} and \ref{eq:vxicorr}) is a delta-symbol, {\em not} a delta-function: 
\begin{equation} \label{eq:dsym}
\delta_{\vc{k}\vc{k}'} = \left\{
\begin{array}{l}
1, \; \vc{k} =    \vc{k}' \\
0, \; \vc{k} \neq \vc{k}' \\
\end{array}
\right.
\end{equation}

It is quite surprising, that such function having a non-zero finite value at one point only produces a non-zero result in integration operations. This can be explained by the fact that the involved integrals have different nature than traditional Riemann integrals, see the next section.


\subsection{Additivity, Gaussianity and convergence} \label{sect:agc}

It is evident, that the ``measure'' $\sqrt{\Delta \vc{k}}$ alone, being not additive, also causes the correspondent integral to diverge. We demonstrate here, that the presence of the random factor $\xi$ cures these discrepancies.

Let us consider an elementary volume $\Delta^{(s)} \vc{k}$ and its subdivision $\{\Delta^{(s)}_{\frac{1}{N},i} \vc{k}\}$ so that\footnote{Here delta with superscript $(s)$ denotes the corresponding set, while in absence of it such delta is a numeric value.}
\begin{equation}
\Delta^{(s)} \vc{k} = \bigcup_{i=1}^N \Delta^{(s)}_{\frac{1}{N},i} \vc{k}
\end{equation}
\begin{equation}
\Delta^{(s)}_{\frac{1}{N},i} \vc{k} \cap \Delta^{(s)}_{\frac{1}{N},j} \vc{k} = \varnothing, \; i \neq j
\end{equation}
\begin{equation}
\Delta_{\frac{1}{N},i} \vc{k} = \Delta_\frac{1}{N} \vc{k} = \frac{1}{N} \Delta \vc{k} 
\end{equation}

Let us assume that the elementary volume $\Delta^{(s)} \vc{k}$ is small enough, that the variations of the factor responsible for power spectrum can be neglected. We set it here to 1 and denote corresponding random measure as $\Phi_1(\cdot)$. 

As the measure is an additive function, it should satisfy the following equation:
\begin{equation} \label{eq:phi_add}
\Phi_1 \left(\Delta^{(s)} \vc{k}\right) = \sum_{i=1}^N \Phi_1 \left(\Delta^{(s)}_{\frac{1}{N},i} \vc{k}\right)
\end{equation}

Following our schema described in \S \ref{sect:sf}, let us define $\Phi_1(\cdot)$ as follows:
\begin{equation} \label{eq:phi1}
\Phi_1 \left(\Delta^{(s)} \vc{k}\right) = \xi \sqrt{\Delta \vc{k}}
\end{equation}
\begin{equation}
\Phi_1 \left(\Delta^{(s)}_{\frac{1}{N},i} \vc{k}\right) = \xi_{\frac{1}{N},i} \sqrt{\Delta_\frac{1}{N} \vc{k}}
\end{equation}
\noindent where
\begin{equation} \label{eq:avg_xi}
\avg{\xi\;\xi^*} = 1
\end{equation}
\begin{equation} \label{eq:avg_xis}
\avg{\xi_{\frac{1}{N},i}\;\xi^*_{\frac{1}{N},j}} = \delta_{ij}
\end{equation} 
Taking into account spectral measure additivity (Eq. \ref{eq:phi_add}), we have the following relation for $\xi$'s:
\begin{equation} \label{eq:sumxi}
\xi = \frac{1}{\sqrt{N}} \sum_{i=1}^N \xi_{\frac{1}{N},i}
\end{equation}

Accounting also for Eq. (\ref{eq:avg_xis}), we see, that Eq. (\ref{eq:avg_xi}) is also true, so our definition of spectral measure Eq. (\ref{eq:phi1}) is self-consistent and such measure is indeed additive. 

Applying Eq. (\ref{eq:sumxi}) recursively and accounting for the central limit theorem, we can also see that additivity requires random variables $\xi$ to be Gaussian\footnote{Requirement Eq. \ref{eq:sphi} is also necessary for this.}.

To demonstrate the convergence mechanism, let us consider the sum of measure elements, corresponding to subdivision  $\{\Delta^{(s)}_{\frac{1}{N},i} \vc{k}\}$:
\begin{equation} \label{eq:conv}
\sum_{i=1}^N \Phi_1 \left(\Delta^{(s)}_{\frac{1}{N},i} \vc{k}\right)
  = \sqrt{\Delta_\frac{1}{N} \vc{k}} \sum_{i=1}^N \xi_{\frac{1}{N},i} 
  = \sqrt{\frac{\Delta \vc{k}}{N}} \sum_{i=1}^N \xi_{\frac{1}{N},i}
  = \xi \sqrt{\Delta \vc{k}}
\end{equation}

We see, that the presence of independent random factors of the same unity variance is necessary here, because the order of their sum cancels with $\sqrt{N}$ in denominator and thus makes the correspondent integral sum converge (if they were coherent, their sum would have the order of $N$, what makes the correspondent integral infinite).

So the convergence mechanism of spectral representations Eq. (\ref{eq:ssr}) and Eq. (\ref{eq:vsr}) has statistical nature and thus is completely different from the one of the inverse Fourier transform, which is a Riemann integral.


\subsection{Relation to the traditional technique}

We can not take a Fourier transform of a stationary field directly: as shown by \cite{T67}, such integral diverges. To do so, we need to introduce a window function $w(\vc{r})$, which cuts out a finite region from the whole space. Then, the correspondent Fourier transform is as follows:
\begin{equation} \label{eq:ftrho}
\begin{array}{ll}
\tilde{\rho}(\vc{k}) 
  & \equiv \dfrac{1}{(2\pi)^3}\displaystyle\int w(\vc{r}) \D{\vc{r}} e^{-i\vc{k}\vc{r}} \rho(\vc{r}) \\
  & =      \dfrac{1}{(2\pi)^3}\displaystyle\int w(\vc{r}) \D{\vc{r}} e^{-i\vc{k}\vc{r}}  
           \int \sqrtd{k'} e^{i\vc{k'}\vc{r}} F(\vc{k}') \, \xi(\vc{k}') \\
  & \simeq  F(\vc{k}) \, \bar{\xi}(\vc{k}) \\
\end{array}
\end{equation}
where 
\begin{equation} \label{eq:xibar}
\bar{\xi}(\vc{k}) = \int \tilde{w}(\vc{k}-\vc{k}') \sqrtd{k'} \, \xi(\vc{k}')
\end{equation}
Let us calculate the auto-correlation of $\bar{\xi}(\vc{k})$:
\begin{equation} \label{eq:xibarcorr}
\begin{array}{ll}
\avg{\bar{\xi}(\vc{k})\; \bar{\xi}^*(\vc{k}')} & =
  \displaystyle\int \tilde{w}  (\vc{k} -\vc{k}_1) \sqrtd{k_1}
  \displaystyle\int \tilde{w}^*(\vc{k}'-\vc{k}_2) \sqrtd{k_2}
  \avg{\xi(\vc{k}_1)\; \xi^*(\vc{k}_2)} \\
  & = \displaystyle\int \tilde{w}(\vc{k}-\vc{k}_1) \tilde{w}^*(\vc{k}'-\vc{k}_1) \D{\vc{k}_1} \\
  & \simeq \delta(\vc{k} -\vc{k}')
\end{array}
\end{equation}
Then, we have the following correlation in Fourier space, if the region cut out by $w(\vc{r})$ is large enough:
\begin{equation} \label{eq:ftrhocorr}
\avg{\tilde{\rho}(\vc{k}) \tilde{\rho}^*(\vc{k}')}
  =  F^2(\vc{k}) \, \delta(\vc{k} -\vc{k}')
\end{equation}
which coincides with the traditional correlation rule, see for example \cite{T67}.

So traditional formalism is a partial limiting case of the one presented here. 

Additionally, we have here direct expressions for scalar and vector random fields Eqs. (\ref{eq:ssr}) and (\ref{eq:vsr}), what is convenient for practical applications. 

It is also convenient, especially for numerical studies, that our formalism does not contain ``magical'' objects such as delta-functions. Indeed, the symbolic rule, suggested by \cite{T67}
\begin{equation} \label{eq:sphi_symb_orig_T67}
\avg{\Phi(d\vc{k})\Phi^*(d\vc{k}')} = \delta(\vc{k} - \vc{k}') F^2(\vc{k}) \D{\vc{k}} \D{\vc{k}'}
\end{equation}
looks much less straightforward than corresponding Eq. (\ref{eq:sphi_symb_orig}) and it does not allow to recover the explicit form of the spectral measure element (Eq. \ref{eq:sphi_symb}). 


\subsection{The problem of direct transform}

Spectral representation Eq. (\ref{eq:ssr}) gives us the inverse transform from the field spectrum. Let us try to find the correspondent direct transform. To determine it let us introduce a normalization factor in Eq. (\ref{eq:ftrho}) so that we get $\delta_{\vc{k}\vc{k}'}$ instead of $\delta(\vc{k} -\vc{k}')$ in Eq. (\ref{eq:xibarcorr}).

If we select the window function as follows
\begin{equation} \label{eq:wtophat}
w(\vc{r}) = \left\{
\begin{array}{l}
1, \; \vc{r} \in \Omega \\
0, \; \vc{r} \notin \Omega \\
\end{array}
\right.
\end{equation}
the normalization factor is $\sqrt{(2\pi)^3/\Omega}$ and we have the following expression for the direct transform of a homogeneous Gaussian field $\rho(\vc{r})$:
\begin{equation} \label{eq:direct}
F(\vc{k}) \, \xi_\Omega(\vc{k})
  = \frac{1}{(2\pi)^{3/2}\sqrt{\Omega}}\int\limits_{\Omega} \D{\vc{r}} e^{-i\vc{k}\vc{r}} \rho(\vc{r})
\end{equation}
(here $\Omega$ is a limited "non-pathological" volume, like sphere or cube).

Unfortunately we can not take a limit with $\Omega \to \mathbb{R}^3$ because the expression on the right side oscillates. 

Indeed, as $\Omega$ is limited, the spectrum on the left side becomes discrete with the value at $\vc{k}$ corresponding to a limited elementary volume $\Delta \vc{k}$. If we increase $\Omega$, the correspondent $\Delta \vc{k}$ gets smaller, and the correspondent random factor $\xi_\Omega$ inside the measure element $\Phi(\Delta \vc{k})$, according to Eq. (\ref{eq:sphi}), gets less dependent on the initial $\xi_\Omega$, and, as a limiting case, is not depending on it at all. Therefore, the limit of Eq. (\ref{eq:direct}) with $\Omega \to \mathbb{R}^3$ does not exist.

Then, the expression Eq. (\ref{eq:direct}) for the limited volume $\Omega$ is the only option we have. However, it seems to be sufficient for practical applications.


\section{Spectral representation of a vector field and related geometrical considerations} \label{sect:vf}

Let us generalize generalize the results obtained for scalar field for the case of a vector one.

\subsection{Spectral representation}

Let us write the correlation function of a homogeneous vector random field $u_i(\vc{r})$ through its power spectrum (here we need to begin with a specific form of the power spectrum, a more general case is considered below in \S \ref{sect:sup}):
\begin{equation} \label{eq:vcf}
C_{ij}(\vc{r}) \equiv \avg{u_i^*(\vc{0})u_j(\vc{r})} = \int e^{i\vc{k}\vc{r}} \mcl{F}_{ij}(d\vc{k}) = \int e^{i\vc{k}\vc{r}} F^2(\vc{k}) T_{ij}(\vch{k}) \D{\vc{k}} 
\end{equation}
\noindent where $\mcl{F}_{ij}(\cdot)$ is a tensor correlation spectral measure, $F^2$ is the scalar part of a power spectrum and $T_{ij}$ is a spectral tensor, satisfying the equations
\begin{equation} \label{eq:vt}
T_{ij}(\vch{k}) = T_{il}(\vch{k})T_{lj}(\vch{k})
\end{equation}
\begin{equation} \label{eq:tsym}
T_{ij}(\vch{k}) = T_{ji}(\vch{k})
\end{equation}

Eqs. (\ref{eq:vt}) and (\ref{eq:tsym}) indicate that $T_{ij}$ is an orthogonal projector, see for example \cite{TB97}.

The spectral representation of the field itself can be written as follows:
\begin{equation} \label{eq:vsr_orig}
u_i(\vc{r}) = \int e^{i\vc{k}\vc{r}} \Phi_i(d\vc{k}) 
\end{equation}
\noindent where $\Phi_i(\cdotp)$ is a complex vector random measure in $\mathbb{R}^3$, satisfying
\begin{equation} \label{eq:vphi}
\avg{\Phi_i(A)\Phi_j^*(B)} = \mcl{F}_{ij}(A\cap B) 
\end{equation}

The correspondent symbolic rule for $\Phi_i(d\vc{k})$ can be written in the following form:
\begin{equation} \label{eq:vphi_symb_orig}
\avg{\Phi_i(d\vc{k})\Phi_j^*(d\vc{k}')} = \delta_{\vc{k}\vc{k}'} \mcl{F}_{ij}(d\vc{k}) =  \delta_{\vc{k}\vc{k}'} F^2(\vc{k}) T_{ij}(\vch{k}) \D{\vc{k}}
\end{equation}

Then the explicit form of the spectral measure element can be represented as follows:
\begin{equation} \label{eq:vphi_symb}
\Phi_i(d\vc{k}) = F(\vc{k}) \, T_{ij}(\vch{k}) \xi_j(\vc{k}) \sqrtd{k},
\end{equation}
\noindent where the complex random field $\xi_i(\vc{k})$ must conform the following correlation rule:
\begin{equation} \label{eq:vxicorr}
\avg{\xi_i(\vc{k})\xi_j^*(\vc{k}')} = \delta_{ij} \delta_{\vc{k}\vc{k}'}.
\end{equation}
\noindent This satisfies Eq. (\ref{eq:vphi_symb_orig}). Finally, Eq. (\ref{eq:vsr_orig}) can be rewritten as follows:
\begin{equation} \label{eq:vsr}
u_i(\vc{r}) = \int e^{i\vc{k}\vc{r}} F(\vc{k}) \,  T_{ij}(\vch{k}) \xi_j(\vc{k}) \sqrtd{k} 
\end{equation}
\noindent where $\xi_i(\vc{k})$ conforms Eq. (\ref{eq:vxicorr}), $F(\vc{k})$ is square root of the scalar part of the power spectrum and the spectral tensor $T_{ij}(\vch{k})$ conforms Eqs. (\ref{eq:vt}) and (\ref{eq:tsym}). 

If the field is real, we also have
\begin{equation} \label{eq:vxisym}
\xi_i(-\vc{k}) = \xi_i^*(\vc{k}).
\end{equation}

A convenient feature of our vector field spectral representation is that the harmonic direction there can be interpreted as a projection, defined by the spectral tensor $T_{ij}$, of a random vector $\xi_i$ having isotropic distribution.

We consider here some consequences of this fact.


\subsection{Spectral tensors} \label{sect:tensor}

Here are the spectral tensors for isotropic fields (potential and solenoidal) and axially symmetrical fields (compressible and Alfv\'enic), as described in \cite{LP12}:
\begin{equation} \label{eq:Tp}
T_{p,ij}(\vch{k}) = \hat{k}_i\hat{k}_j
\end{equation}
\begin{equation} \label{eq:Ts}
T_{s,ij}(\vch{k}) = \delta_{ij}-\hat{k}_i\hat{k}_j
\end{equation}
\begin{equation} \label{eq:Tc}
T_{c,ij}(\vch{k}) = \frac{
(\vch{k}\vch{\lambda})^2\hat{k}_i\hat{k}_j+\hat{\lambda}_i\hat{\lambda}_j-(\vch{k}\vch{\lambda})(\hat{k}_i\hat{\lambda}_j+\hat{\lambda}_i\hat{k}_j)
}{1-(\vch{k}\vch{\lambda})^2}
\end{equation}
\begin{equation} \label{eq:Ta}
T_{a,ij}(\vch{k}) = \delta_{ij}-\hat{k}_i\hat{k}_j - \frac{
(\vch{k}\vch{\lambda})^2\hat{k}_i\hat{k}_j+\hat{\lambda}_i\hat{\lambda}_j-(\vch{k}\vch{\lambda})(\hat{k}_i\hat{\lambda}_j+\hat{\lambda}_i\hat{k}_j)
}{1-(\vch{k}\vch{\lambda})^2}
\end{equation}
\noindent where $\vch{\lambda}$ defines the symmetry axis.

As noted before, these tensors are orthogonal projectors. In the following section we consider the ranges of them.


\subsection{Ranges} \label{sect:range}

Let us choose the coordinate system so that $\vch{\lambda}$  and $\vch{k}$ lay in $zx$ plane:
\begin{equation} \label{eq:lambda}
\vch{\lambda} = (\sin\theta_\lambda, 0, \cos\theta_\lambda)^T
\end{equation}
\begin{equation} \label{eq:k}
\vch{k} = (\sin\theta, 0, \cos\theta)^T
\end{equation}

Then the corresponding tensors and their ranges are as follows (dependence on $\theta_\lambda$ cancels here):

potential:
\begin{equation} \label{eq:Tp_}
T_p = 
\begin{pmatrix}
         \sin^2\theta & 0 &  \cos\theta\sin\theta \\
                    0 & 0 &                     0 \\
 \cos\theta\sin\theta & 0 &          \cos^2\theta
\end{pmatrix}
\end{equation}
\begin{equation} \label{eq:Tp_r}
\mathrm{range}\,T_p = \left\{(\sin\theta, 0, \cos\theta)^T\right\}
\end{equation}

solenoidal:
\begin{equation} \label{eq:Ts_}
T_s = 
\begin{pmatrix}
         \cos^2\theta & 0 & -\cos\theta\sin\theta \\
                    0 & 1 &                     0 \\
-\cos\theta\sin\theta & 0 &          \sin^2\theta
\end{pmatrix}
\end{equation}
\begin{equation} \label{eq:Ts_r}
\mathrm{range}\,T_s = \left\{(-\cos\theta, 0, \sin\theta)^T,\, (0, 1, 0)^T\right\}
\end{equation}

compressible:
\begin{equation} \label{eq:Tc_}
T_c = 
\begin{pmatrix}
         \cos^2\theta & 0 & -\cos\theta\sin\theta \\
                    0 & 0 &                     0 \\
-\cos\theta\sin\theta & 0 &          \sin^2\theta
\end{pmatrix}
\end{equation}
\begin{equation} \label{eq:Tc_r}
\mathrm{range}\,T_c = \left\{(-\cos\theta, 0, \sin\theta)^T\right\}
\end{equation}

Alfv\'enic:
\begin{equation} \label{eq:Ta_}
T_a = 
\begin{pmatrix}
0 & 0 & 0 \\
0 & 1 & 0 \\
0 & 0 & 0
\end{pmatrix}
\end{equation}
\begin{equation} \label{eq:Ta_r}
\mathrm{range}\,T_a = \left\{(0, 1, 0)^T\right\}
\end{equation}

Geometrical interpretation of these ranges is given in the next section.



\subsection{Component separation} \label{sect:sep}

Let us consider an arbitrary vector field, not necessarily solenoidal.

As we can see from \S \ref{sect:range}, $T_\alpha$ has the following geometrical meaning: for potential field it projects a vector to the line, defined by $\vch{k}$, for compressible field it projects it to the line perpendicular to $\vch{k}$ in the plane containing $\vch{k}$ and $\vch{\lambda}$ and for Alfv\'enic field it projects it to the line perpendicular to both $\vch{k}$ and $\vch{\lambda}$.

So the vectors, defining the one-dimensional ranges of potential, compressible and Alfv\'enic tensors form the local orthonormal basis $\{\vch{e}_p(\vch{k}),\vch{e}_c(\vch{k},\vch{\lambda}),\vch{e}_a(\vch{k},\vch{\lambda})\}$. This means, that if we also account for Eqs. \ref{eq:vt} and \ref{eq:tsym}, the following is true:
\begin{equation} \label{eq:ort}
T_{\alpha,ij}(\vch{k}) T_{\beta,jl}(\vch{k}) = \delta_{\alpha\beta} T_{\alpha,il}(\vch{k})
\end{equation}
Here $\alpha,\beta \in \{p,c,a\}$ and summation over repeating Greek indices is not evaluated.

Therefore, if we have the total field harmonic (Eq. \ref{eq:ort} allows us to use the same 3-component $\xi$ for all of 3 components $\{p,c,a\}$ so that Khinchin–Kolmogorov theorem holds)
\begin{equation} \label{eq:Fu}
\tilde{u}_i(\vc{k}) = \sum_\alpha F_\alpha(\vc{k}) \,  T_{\alpha,ij}(\vch{k}) \xi_j(\vc{k})
\end{equation}
we can separate it to components by applying the corresponding spectral tensors as follows:
\begin{equation} \label{eq:Fu_sep}
\tilde{u}_{\alpha,i}(\vc{k}) = T_{\alpha,ij}(\vch{k}) \tilde{u}_j(\vc{k})
\end{equation}


\subsection{Superposition} \label{sect:sup}

Using Eq. (\ref{eq:Fu_sep}), any vector field can be decomposed to the components $\{p,c,a\}$ in Fourier space. If, additionally, its harmonics are delta-correlated (i.e. if the field is Gaussian), superposition of these components can be represented as follows:
\begin{equation} \label{eq:sup1}
u_i(\vc{r}) = \sum_\alpha \int e^{i\vc{k}\vc{r}} F_\alpha(\vc{k}) \,  T_{\alpha, ij}(\vch{k}) \xi_j(\vc{k}) \sqrtd{k} 
\end{equation}

Accounting for Eqs. (\ref{eq:vt}), (\ref{eq:tsym}),  (\ref{eq:vxicorr}) and (\ref{eq:ort}) its correlation function can be written in the following way:
\begin{equation} \label{eq:corrsup1}
C_{ij}(\vc{r}) = \int e^{i\vc{k}\vc{r}} \left( \sum_\alpha F_\alpha^2(\vc{k}) \, T_{\alpha,ij}(\vch{k}) \right) \D{\vc{k}} 
\end{equation}
where $\alpha \in \{p,c,a\}$.

Here unity vector $\vch{\lambda}$ contained in the expressions for $T_{\alpha,ij}$ can be any, but for practical applications it should be set to the real symmetry axis of the field power spectrum.


\subsection{Power spectrum component separation} \label{sect:pssep}

As found in \S \ref{sect:sup}, the power spectrum of a Gaussian field can be represented as follows:
\begin{equation} \label{eq:ps}
F_{ij}(\vc{k}) = \sum_\alpha F_\alpha^2(\vc{k}) \, T_{\alpha,ij}(\vch{k}) 
\end{equation}
where $\alpha \in \{p,c,a\}$. Then, accounting for Eq. (\ref{eq:ort}), we can separate it to components:
\begin{equation} \label{eq:pssep}
F_{\alpha,ij}(\vc{k}) = T_{\alpha,il}(\vch{k}) \, F_{lj}(\vc{k}) = F_\alpha^2(\vc{k}) \, T_{\alpha,ij}(\vch{k}) 
\end{equation}
and, as $\mathrm{tr} \, {\bf T}_\alpha = 1$, taking trace of the result we can extract the scalar part of the power spectrum component:
\begin{equation} \label{eq:pssp}
F_\alpha^2(\vc{k}) = F_{\alpha,ii}(\vc{k}) 
\end{equation}


\section{Some practical applications related to vector field}

\subsection{Restoring $z$-component of a solenoidal field} \label{sect:z}

Solution of this problem may be useful for restoring of 3D magnetic field using the technique of Faraday depolarization tomography, introduced by \cite{LY18}.

Suppose we have only $x$ and $y$ components of a solenoidal field and try to restore its $z$-component. 

Let us set $z$-component of such field $u_i(\vc{r})$ to zero and calculate its harmonic $\tilde{u}_i(\vc{k})$, which therefore has a zero $z$-component too. Then, to find the harmonic's missing $z$-component $a(\vc{k})$ we can demand, that potential component of the combined field is zero:

\begin{equation} \label{eq:s1}
T_{p,ij}(\vch{k})(\tilde{u}_j(\vc{k})+a(\vc{k})\hat{e}_{z,j}) = 0_i
\end{equation}

This is possible, because non-zero results of applying of $T_p$ to any vectors are collinear to each other and to $\vch{k}$, see \S \ref{sect:range}. Then we have the following expression for the missing $z$-component:

\begin{equation} \label{eq:s2}
a(\vc{k}) = -\frac{\hat{k}_i T_{p,ij}(\vch{k})\tilde{u}_j(\vc{k})}{\hat{k}_i T_{p,ij}(\vch{k})\hat{e}_{z,j}}
\end{equation}

However, this expression has uncertainty $0/0$ at $k_z=0$ so harmonics corresponding to zero $k_z$ cannot be restored. Therefore, to find magnetic field in real space, we need additional information, namely z-component of magnetic field integrated over line of sight, which could be obtained for example from Faraday rotation data.

If we are interested in component separation only, it is possible to find $\vch{\lambda}$ taking into account known scalar factors of MHD modes spectra (see \citealp{LP12}) and do component separation for known harmonics.


\subsection{Restoring amplitudes of a POS-projected magnetic field from its directions} \label{sect:dir}

The technique of Synchrotron Polarization Gradients, introduced by \cite{LY18}, allows us to calculate directions of POS-projected magnetic field only. On the other hand, if the extent of LOS-integrated layer is large enough, the {\em projected} magnetic field can be considered solenoidal, what can be used to recover unknown magnetic field amplitudes.

Let us write the magnetic field harmonic as a discrete Fourier transform (all vectors are 2D here):
\begin{equation}
\vc{h}_i = \sum_j a_j \vc{d}_j \; e^{-i\vc{k}_i\vc{r}_j}
\end{equation}
where $\vc{d}_j$ and $a_j$ are direction and unknown amplitude of magnetic field at radius vector $\vc{r}_j$.

Potential component of the harmonic can be written as follows:
\begin{equation}
\vc{h}_{p,i} = \tpki \cdot \vc{h}_i = \sum_j a_j \tpki \cdot \vc{d}_j \; e^{-i\vc{k}_i\vc{r}_j}
\end{equation}

We would like to minimize the potential component contribution. Then, the correspondent function to minimize is as follows:
\begin{equation}
L = \sum_i \vc{h}_{p,i} \vc{h}_{p,i}^* + \left(\sum_i a_i - n \right)^2
  = \sum_j \sum_{j'} a_j a_{j'} C_{jj'} + \left(\sum_i a_i - n \right)^2
\end{equation}
where the second term is introduced to exclude trivial solution $a_j \equiv 0$, $n$ is the total number of points and $ C_{jj'}$ is as follows:
\begin{equation}
  C_{jj'} = \sum_i (\tpki \cdot \vc{d}_j) \cdot (\tpki \cdot \vc{d}_{j'})\; \cos \vc{k}_i(\vc{r}_j - \vc{r}_{j'})
\end{equation}

Then, for the minimum over $a_k$ we have:
\begin{equation}
\frac{\partial L}{\partial a_k} = 2\sum_j a_j C_{jk} + 2\sum_j a_j - 2n = 0  
\end{equation}
so to find amplitudes $a_j$ we need to solve the following system of $n$ linear equations:
\begin{equation}
\sum_j a_j\; (C_{jk} + 1) = n  
\end{equation}

These equations allow to restore field amplitudes in absence of noise, but adding even a small white noise to original field results in complete loss of its correlation with the restored field. The reason of it is that solenoidality in this case is achieved at a cost of unrealistically high dispersion of amplitudes. 

To suppress this tendency we can modify $L$ so that the second term contains the amplitude dispersion:
\begin{equation}
L = \sum_j \sum_{j'} a_j a_{j'} C_{jj'} + \gamma \sum_i (a_i - 1)^2
\end{equation}
where $\gamma$ is some adjustment coefficient.

Then, the correspondent system of equations for $a_j$ is as follows:
\begin{equation}
\sum_j a_j\; (C_{jk} + \gamma \, \delta_{jk}) = \gamma  
\end{equation}
where the behavior of optimal $\gamma$ can be studied using numerical simulations.

\subsection{Solenoidal field component separation in 2D} \label{sect:sep2}

Let us consider a slice $U_i(\vc{R})$ of a 3D solenoidal field $u_i(\vc{r})$, cut by the LOS window function $w(z)$, within a 2D area, defined by the POS window function $w(\vc{R})$. Then its 2D Fourier transform of its projection over LOS is as follows:   

\begin{equation} \label{eq:U}
\begin{array}{ll}
\tilde{U}_i(\vc{K}) 
  & \equiv \frac{1}{(2\pi)^2}\int w(\vc{R}) \D{\vc{R}} e^{-i\vc{K}\vc{R}} \int w(z) \D{z} u_i(\vc{r}) \\
  & =      \frac{1}{(2\pi)^2}\int w(\vc{R}) \D{\vc{R}} e^{-i\vc{K}\vc{R}} \int w(z) \D{z} 
           \int \sqrtd{k'} e^{i\vc{k'}\vc{r}} F(\vc{k}') \, T_{ij}(\vch{k}') \, \xi_j(\vc{k}') \\
  & \simeq 2\pi \int \tilde{w}^*(k_z) \sqrt{dk_z} F(\vc{k}) \, T_{ij}(\vch{k}) \, \bar{\xi}_j(\vc{k}) \\
\end{array}
\end{equation}
where 
\begin{equation} \label{eq:2s2}
\bar{\xi}_i(\vc{K},k_z') = \int \sqrtd{K'} \tilde{w}(\vc{K}-\vc{K}') \, \xi_i(\vc{K}', k_z)
\end{equation}

Below we need the following correlation:
\begin{equation} \label{eq:2s3}
\avg{\bar{\xi}_i(\vc{K},k_z) \bar{\xi}^*_j(\vc{K},k_z')} = \frac{\Omega}{(2\pi)^2} \, \delta_{k_z k_z'} \delta_{ij}
\end{equation}
\noindent where $\Omega \equiv \int w^2(\vc{R}) \D{\vc{R}}$. 

Let us calculate the variance of $\tilde{U}_i$:
\begin{equation} \label{eq:2s4}
\begin{array}{ll}
\avg{\tilde{U}_i(\vc{K}) \tilde{U}^*_j(\vc{K})} 
  & \simeq (2\pi)^2 
    \int \tilde{w}^*(k_z)  \sqrt{dk_z}  F(\vc{K},k_z)  T_{ik}(\vc{K},k_z)  \, 
    \int \tilde{w}(k_z')   \sqrt{dk_z'} F(\vc{K},k_z') T_{jl}(\vc{K},k_z') \, 
    \avg{\bar{\xi}_k(\vc{K},k_z) \, \bar{\xi}^*_l(\vc{K},k_z')} \\
  & = \Omega \int |\tilde{w}(k_z)|^2  \D{k_z}  F^2(\vc{k}) \, T_{ij}(\vch{k}) \\
\end{array}
\end{equation}

Let us assume $\vc{K}\parallel\vc{\Lambda}\parallel\vch{e}_x$, where $\vc{\Lambda}$ is a POS projection of $\vch{\lambda}$. Then the variances of compressible and Alfv\'enic components are as follows:
\begin{equation} \label{eq:2s5}
\sigma^2_{\alpha,ij}(K_x) \equiv \avg{\tilde{U}_{\alpha,i}(K_x) \tilde{U}^*_{\alpha,j}(K_x)}  
  = \Omega \int |\tilde{w}(k_z)|^2 \D{k_z} F^2_\alpha(K_x,k_z) \, T_{\alpha,ij}(\theta) 
\end{equation}
where $i,j\in\{x,y\}$, $\alpha\in\{c,a\}$ and $T_\alpha$ is presented in \S \ref{sect:range}. The $x,y$ components of $T_\alpha$ are zero with the exception of
\begin{equation} \label{eq:2s6}
T_{c,xx} = \cos^2 \theta = \frac{k_z^2}{K_x^2 + k_z^2}
\end{equation}
\begin{equation} \label{eq:2s7}
T_{a,yy} = 1
\end{equation}

This allows us to separate the components\footnote{When processing real data we can replace ensemble averaging with averaging over some finite ranges of $K_x$.}: the variance of the harmonic parallel to $\vc{\Lambda}$ gives us the compressible component:
\begin{equation} \label{eq:sigma_c}
\sigma^2_c(K_x) \equiv \avg{\tilde{U}_{x}(K_x) \tilde{U}^*_{x}(K_x)}  
  = \Omega \int \frac{k_z^2 |\tilde{w}(k_z)|^2}{K_x^2 + k_z^2} \D{k_z} F^2_c(K_x,k_z) 
\end{equation}
and the variance of the harmonic perpendicular to $\vc{\Lambda}$ gives us the Alfv\'enic component:
\begin{equation} \label{eq:sigma_a}
\sigma^2_a(K_x) \equiv \avg{\tilde{U}_{y}(K_x) \tilde{U}^*_{y}(K_x)}  
  = \Omega \int |\tilde{w}(k_z)|^2 \D{k_z} F^2_a(K_x,k_z) 
\end{equation}

To be able to compare these values we need to account for difference in weights, with which the spectra are integrated over $k_z$. We can do this by introducing the normalized values
\begin{equation} \label{eq:2s10}
\hat{\sigma}^2_c(K_x) \equiv N^2_c \sigma^2_c(K_x) 
\end{equation}
\begin{equation} \label{eq:2s11}
\hat{\sigma}^2_a(K_x) \equiv N^2_a \sigma^2_a(K_x) 
\end{equation}
where
\begin{equation} \label{eq:2s12}
N^2_c \equiv \left(\int \frac{k_z^2 |\tilde{w}(k_z)|^2}{K_x^2 + k_z^2} \D{k_z}\right)^{-1} 
\end{equation}
\begin{equation} \label{eq:2s13}
N^2_a \equiv \left(\int |\tilde{w}(k_z)|^2 \D{k_z}\right)^{-1} 
\end{equation}
For example, if we take
\begin{equation} \label{eq:2s14}
w(z) = \left\{
\begin{array}{l}
1, \; |z| \leqslant \frac{\Delta z}{2}\\
0, \; |z| > \frac{\Delta z}{2}\\
\end{array}
\right. 
\end{equation}
we have
\begin{equation} \label{eq:2s15}
N^2_c = \pi K_x \left(1 + \coth \frac{K_x \Delta z}{2} \right)
\end{equation}
\begin{equation} \label{eq:2s16}
N^2_a = \frac{2\pi}{\Delta z} 
\end{equation}

There is an alternative way of obtaining Eqs. \ref{eq:sigma_c} and \ref{eq:sigma_a}. Taking into account the geometrical properties of $T_\alpha$ (see \S \ref{sect:sep}) we can write the following expressions for compressible and Alfv\'enic 2D harmonics:

\begin{equation} \label{eq:2sUcx}
\tilde{U}_{c,x}(K_x) = \int \tilde{w}^*(k_z) \sqrt{d k_z} \frac{k_z}{\sqrt{K_x^2 + k_z^2}} F_c(K_x,k_z) \, \bar{\xi}_c(K_x,k_z) 
\end{equation}
\begin{equation} \label{eq:2s18}
\tilde{U}_{c,y}(K_x) = 0  
\end{equation}
\begin{equation} \label{eq:2sUay}
\tilde{U}_{a,y}(K_x) = \int \tilde{w}^*(k_z) \sqrt{d k_z} F_a(K_x,k_z) \, \bar{\xi}_a(K_x,k_z) 
\end{equation}
\begin{equation} \label{eq:2s20}
\tilde{U}_{a,x}(K_x) = 0  
\end{equation}
where
\begin{equation} \label{eq:2s21}
\bar{\xi}_\alpha(\vc{k}) = \hat{e}_{\alpha,i}(\vch{k}) \, T_{\alpha,ij}(\vch{k}) \, \bar{\xi}_j(\vc{k}) \\
\end{equation}

Using Eqs. \ref{eq:2sUcx} and \ref{eq:2sUay} we can derive the variances as in Eqs. \ref{eq:sigma_c} and \ref{eq:sigma_a}.



\end{document}